\documentclass[journal,12pt,onecolumn,draftclsnofoot]{IEEEtran}
\usepackage{blindtext}
\usepackage{graphicx}
\usepackage{float}
\usepackage{cite}
\usepackage{amsmath}
\usepackage{amssymb}
\usepackage{epstopdf}
\usepackage{makecell}
\usepackage{lipsum}
\usepackage{mathrsfs}
\usepackage{url}
\usepackage{xcolor}
\usepackage{algorithm}
\usepackage{algorithmic}
\usepackage{amsmath}
\usepackage{lipsum}
\usepackage{enumitem}
\usepackage[utf8]{inputenc}
\usepackage[font=footnotesize,caption=false,subrefformat=parens,labelformat=parens]{subfig}

\title{Channel Rank Improvement in Urban Drone Corridors Using Passive Intelligent Reflectors}
\author{Ender~Ozturk,~Chethan~Kumar~Anjinappa,~Fatih~Erden,~\.{I}smail G\"{u}ven\c{c}, Huaiyu~Dai,~and~Arupjyoti Bhuyan
\thanks{This work is supported in part by the INL Laboratory Directed Research Development (LDRD) Program under DOE Idaho Operations Office Contract DEAC07-05ID14517, and Army Research Office (ARO) under Grant W911NF-16-1-0448. 
E. Ozturk, C. K. Anjinappa, F. Erden, \.{I}.  G\"{u}ven\c{c}, and H. Dai are with the Dept. of Electrical and Computer Eng., NC State Univ., Raleigh, NC 27606 (e-mail:\{eozturk2,canjina,ferden,iguvenc,hdai\}@ncsu.edu). 
A. Bhuyan is with the Idaho National Laboratory, Idaho Falls, ID 83415 (e-mail:arupjyoti.bhuyan@inl.gov).}

}

\begin{document}

\maketitle
\begin{abstract}
Multiple-input multiple-output (MIMO) techniques can help in scaling the achievable air-to-ground (A2G) channel capacity while communicating with drones. However, spatial multiplexing with drones suffers from rank deficient channels due to the unobstructed line-of-sight~(LoS), especially in millimeter-wave~(mmWave) frequencies that use narrow beams. One possible solution is utilizing low-cost and low-complexity metamaterial-based intelligent reflecting surfaces~(IRS) to enrich the multipath environment, taking into account that the drones are restricted to fly only within well-defined drone corridors. A hurdle with this solution is placing the IRSs optimally. In this study, we propose an approach for IRS placement with a goal to improve the spatial multiplexing gains, and hence to maximize the average channel capacity in a predefined drone corridor. Our results at 6~GHz, 28~GHz and 60~GHz show that the proposed approach increases the average rates for all frequency bands for a given drone corridor, when compared with the environment where there are no IRSs present, and IRS-aided channels perform close to each other at sub-6 and mmWave bands.
\end{abstract}

\begin{IEEEkeywords}
Drone corridor, Intelligent Reflecting Surfaces~(IRS), metasurfaces, MIMO communications, mmWave.
\end{IEEEkeywords}
\IEEEpeerreviewmaketitle
\section{Introduction}

\IEEEPARstart{M}{illimeter-wave}~(mmWave) bands offer abundant amount of free spectrum to be exploited for increasing the mobile broadband data rates in 5G networks and beyond. The trade-off is the higher path loss that dictates a larger number of base stations~(BSs) to achieve a similar coverage performance with a sub-6~GHz wireless network. Particularly in an urban drone corridor~\cite{UAM_FAA_ConOps,bhuyan2019secure}, BS deployments around the corridor~\cite{singh_VTC_2021} introduce dominant line-of-sight~(LoS) paths throughout the aerial service area~(ASA). This diminishes the spatial multiplexing gains for 
multiple-input multiple-output~(MIMO) communication links with the drones. 
One way of enhancing the channel capacity via spatial multiplexing is by deploying reflectors with beamsteering and focusing functions. This will help create a favorable propagation environment by introducing new multipath components (MPCs) to the channel, hence, increasing the overall spatial diversity.

Optimal placement of BSs has been studied with several different performance criteria in the literature, such as cost, coverage, and capacity~\cite{chethans10,chechatns13}. 

Optimal placement of reflectors is also studied with similar motivations. In~\cite{anjinappa2020base}, authors maximize coverage area considering BSs and passive metallic reflectors~(PMRs) in an outdoor urban setting, while~\cite{wahab_passive_coverage} presents channel sounding results to explore coverage gains when using passive reflectors. 
Metallic reflectors do not have the capability to tilt the incident wave, but they only reflect towards the specular direction without focusing, which yields excess loss in comparison with metamaterial-based intelligent reflecting surfaces~(IRSs) with beamforming capabilities~\cite{beamforming_survey}. The necessity to have a certain physical tilt angle also makes the PMRs less practical in an urban area due to cosmetic and logistic reasons. 
Optimal network design using IRSs is a relatively new topic in the literature~\cite{contemp_survey}. In~\cite{B_LoS_coverage_mmWave}, optimum placement of IRSs is studied to eliminate the coverage holes in an urban area, which shows that properly deployed IRSs can fulfill the predefined coverage goals even with imperfect context information. In~\cite{ozcan2020reconfigurable},  IRS placement problem is formulated by considering their sizes and operating modes to increase the reliability of vehicle-to-everything (V2X) communications. 

It is already shown in~\cite{RankImprovement_IRS} that a single IRS can improve the channel matrix rank to support spatial multiplexing; however, optimal placement of IRSs remains unaddressed. In this work, we study the optimal placement of IRSs to maximize the average channel capacity in a drone corridor. 
For a given set of candidate IRS locations, a BS, and a drone corridor (defined in terms of ASAs), and assuming MIMO communication, we assign IRSs to aerial surface areas such that the average downlink (DL) capacity in the corridor is maximized. 
Optimal placement of IRSs is studied in a similar way to our study in~\cite{RankImprovement_IRS}, where the optimal location of a single IRS is sought to increase the channel capacity with spatial multiplexing. In our work, we used path loss models given in the literature to calculate the rates. Subsequently, we formulated and solved an optimization problem to find optimal locations of passive IRSs in a given candidate set, to maximize the average capacity in 
the ASAs (drone corridor), which to our knowledge has not been explored in the literature.

\section{System Model} \label{system_model}
\begin{figure}
\vspace{-2mm}
\centerline{\includegraphics[width=\columnwidth]{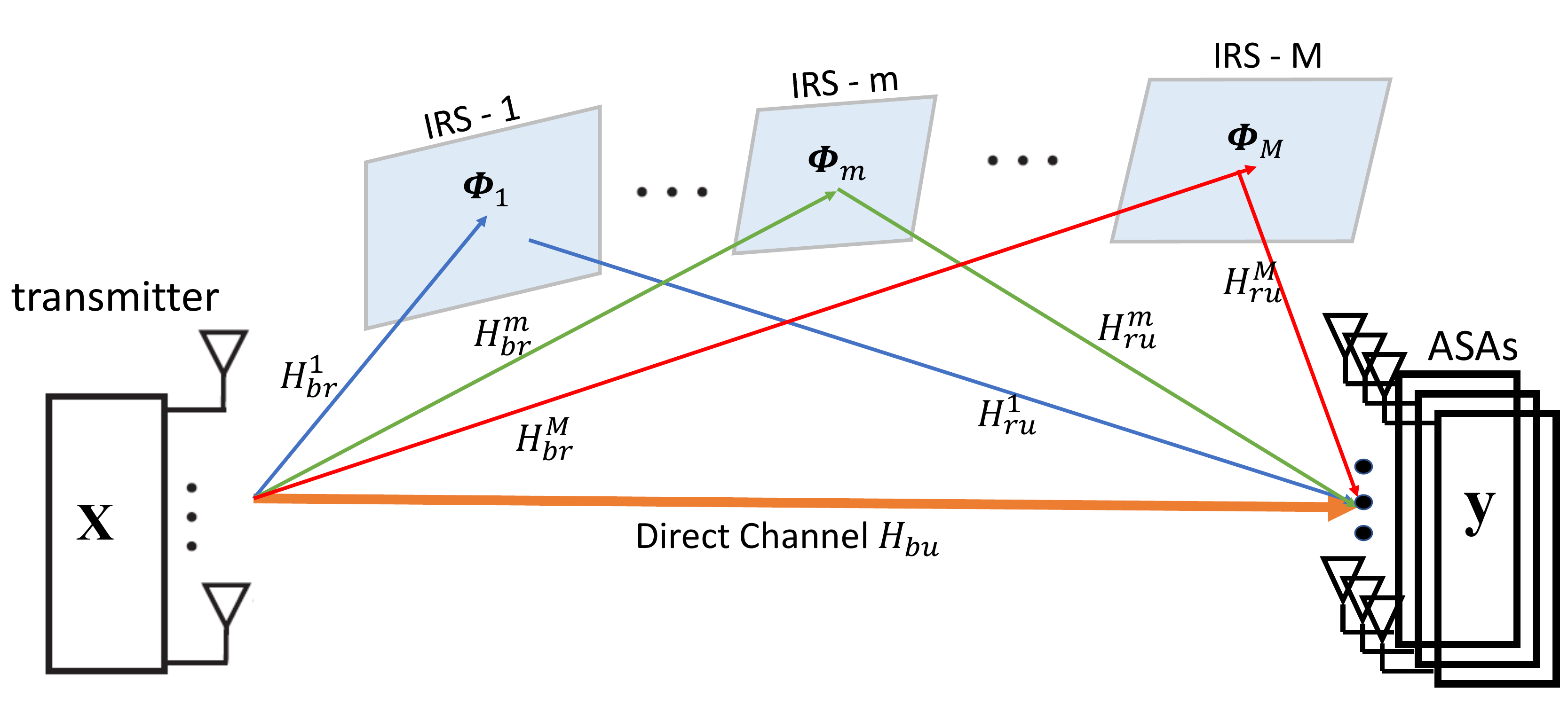}}\vspace{-2mm}
\caption{System model: introducing multiple IRSs in the environment can improve the spatial multiplexing gains at given locations in a drone corridor.} 

\label{Fig:system_model}
\vspace{-2mm}
\end{figure}

We consider an IRS-assisted MIMO system as in Fig.~\ref{Fig:system_model}, where the users are defined to be within equally spaced ASAs that constitute a drone corridor. In this study, only passive reflectors are considered that do not require any dynamic beamforming, and hence have cost and power-efficiency advantages. In this letter, aligned with some other existing literature~\cite{passive_1, passive_2}, we still refer to such reflectors as IRSs, as they are designed to have unique reflection characteristics using meta-surfaces. Consequently, in this design,  a single IRS is assumed to serve only a single ASA. 
Let the channel be an $N_{\rm rx} \times N_{\rm tx}$ MIMO channel. The total number of ASAs, and IRSs and unit cells on the $m^{\rm{th}}$ IRS are given as $\xi$, $\psi$, and $N_m$, respectively. Channel matrix can be decomposed into two components, namely, line-of-sight~(LoS) component, and guided reflection component.

\subsection{LoS Transmission}
The received signal is the superposition of LoS component and the guided reflection component in this model. The received LoS signal $\textbf{y}_{\text{LoS}} \in \mathbb{C}^{N_{\rm rx} \times 1}$ is defined as $\textbf{y}_{\text{LoS}} = \textbf{H}_{\rm bu} \textbf{V}\textbf{x}+\textbf{n}$,
where $\textbf{x} \in \mathbb{C}^{N_{\rm tx} \times 1}$ is the transmitted signal, $\textbf{V} \in \mathbb{C}^{N_{\rm tx} \times N_{\rm tx}}$ is the precoding matrix at the BS, $\textbf{n}~\sim\mathcal{CN}(0,\sigma^{2}\textbf{I})$ is the complex additive white Gaussian noise (AWGN), and $\textbf{H}_{\rm bu} \in \mathbb{C}^{N_{\rm rx} \times N_{\rm tx}}$ is the direct channel. The singular value decomposition of $\textbf{H}_{\rm bu}$ is equal to $\textbf{U}\mathbf{\Lambda}\textbf{V}^{H}$, where $\textbf{U} \in \mathbb{C}^{N_{\rm rx} \times N_{\rm rx}}$ and $\textbf{V}$ are unitary matrices by definition, and $\mathbf{\Lambda} \in \mathbb{R}_+^{N_{\rm rx} \times N_{\rm tx}}$ is the diagonal singular value matrix. 

Given these, the received signal at the user side can be rewritten as $ \textbf{U}^H\textbf{y}_{\text{LoS}} =  \textbf{U}^H \textbf{H}_{\rm bu} \textbf{V} \textbf{x} + \textbf{U}^H \textbf{n}$. 
A single entry of the channel matrix, $\textbf{H}_{\rm bu}$, is given by $\sqrt{\beta^{t,u}}e^{j2\pi \frac{d^{t,u}}{\lambda_c}}$, where $\beta^{t,u}$ is the free-space path loss component, $d^{t,u}$ is the distance between the $t^{\text{th}}$ transmitter antenna and the $u^{\text{th}}$ receiver antenna, and $\lambda_c$ is the wavelength. We assume $\beta^{t,u}$ to be approximately equal for all $(t,u)$ pairs in the far field. In this LoS scenario, $\textbf{H}_\text{bu}$ has a single non-zero singular value, $\lambda_{d}$, hence, the channel capacity in bits/s/Hz is
\begin{equation}
    R_{\text{LoS}} = \text{log}_2 \left( 1+\frac{P_{\text{tot}}\lambda_{d}^2}{\sigma^2}  \right),
\end{equation}
where $P_\text{tot}$ is the total radiated power and $\sigma^2$ is the 
noise power. 

\subsection{Guided Reflection Component}
When IRSs are used, new received components enrich the multipath environment and enable additional data streams to increase the capacity. 
Let $\textbf{H}_{\rm br}^m \in \mathbb{C}^{N_m \times N_{\rm tx}}$ be the channel between the transmit antennas and the $m^{\text{th}}$ IRS. A single normalized matrix element that associates the $i^{\text{th}}$ cell of the $m^{\text{th}}$ reflector to the $t^{\text{th}}$ transmit antenna $\hat{h}_{\rm br}^{t,m,i}$ is expressed as
\begin{equation}
    \hat{h}_{\rm br}^{t,m,i} = e^{j2\pi d_{\rm br}^{t,m,i}\frac{1}{\lambda_c}}, 
\end{equation}
where $d_{\rm br}^{t,m,i}$ is the distance between the transmit antenna and the unit cell under consideration.

Now let $\textbf{H}_{\rm ru}^m \in \mathbb{C}^{N_{\rm rx} \times N_m}$ be the channel between the $m^{\text{th}}$ IRS and the receive antennas. Similarly, a single normalized matrix element that associates the $u^{\text{th}}$ receive antenna to $i^{\text{th}}$ cell of the $m^{\text{th}}$ reflector $\hat{h}_{\rm ru}^{u,m,i}$ is expressed as $\hat{h}_{\rm ru}^{u,m,i} = e^{j2\pi \frac{d_{\text{ru}}^{u,m,i}}{\lambda_c}},$
where $d_{\text{ru}}^{u,m,i}$ is the distance between the receive antenna and the unit cell under consideration.  
Given these definitions, the channel equation for the guided reflection scenario becomes $\textbf{y}_{\text{u}} = \textbf{H}_{\rm c} \textbf{V}_{\text{u}}\textbf{x}+\textbf{n},$
where $\textbf{V}_{\text{u}}$ is the precoding matrix at the BS, and $\textbf{H}_{\rm c}$ is the compound channel matrix, which can be expressed in terms of the set $\mathcal{L}$ that contains the indexes of IRSs that are dedicated to serve a particular ASA. It can be expressed as:
\begin{equation}\label{compound_channel}
    \textbf{H}_{\rm c} = \left( \sum_{m \in \mathcal{L}} \textbf{H}_{\rm ru}^m\mathbf{\Phi}^m\textbf{H}_{\rm br}^m\right)+\textbf{H}_{\rm bu},
\end{equation}
where $\mathbf{\Phi}^m = \alpha_m \text{diag}(e^{j\phi_1^m},e^{j\phi_2^m},\ldots,e^{j\phi_{N_m}^m})$ 
is the diagonal phase contribution matrix of the $m^{\text{th}}$ reflector, and $\alpha_m$ is the reflection loss. The expression in parenthesis in (\ref{compound_channel}) includes the free-space path loss from the transmitter to the receiver via the $m^{\text{th}}$ reflector, $\beta_{c}^m$. In this study, the path loss model in~\cite{bjornson_physics} is used, which can be modified for a square shape reflector as
\begin{equation}
    \beta_c^m = \frac{G_t G_u}{(4\pi)^2}\left(\frac{N_m {d_{\lambda}^{m}}^2 }{d_{\rm br}^m d_{\rm ru}^m}\right)^2 \cos^2(\phi_m^t),\label{Eq_cosine}
\end{equation}
where $G_t$ and $G_u$ are the antenna gains of the transmitter and the receiver, $d_{\lambda}^{m}$ is the size of the unit cell of $m^{\text{th}}$ reflector, and $\phi_m^t$ is the angle of incidence.

Let the total number of non-zero singular values of the compound channel be $S$, where each singular value $\lambda_s$ is a function of physical positions of all reflectors in $\mathcal{L}$ as well as their orientations, sizes, and phase contributions. 
Then, the total rate of the compound channel is given in~\cite{tse} as
\begin{equation}
    R_{\rm IRS} =\sum_{s=1}^{S} \text{log}_2 \left( 1+\frac{P_{\text{s}}\lambda_s^2}{\sigma^2}  \right),\label{Eq_ChanRate}
\end{equation}
where $P_{\text{s}}$ is the amount of power allocated for a specific data stream, and is defined using water filling algorithm given by
$P_s = \left(\mu-\frac{\sigma^2}{\lambda_s^2} \right)^+$, 
where $\mu$ is chosen to satisfy the total power constraint, i.e., $\sum_s P_s = P_{\text{tot}}$.

\section{Drone Corridor Reflector Placement}\label{optimization}

\begin{figure}
\centerline{\includegraphics[width=0.8\columnwidth]{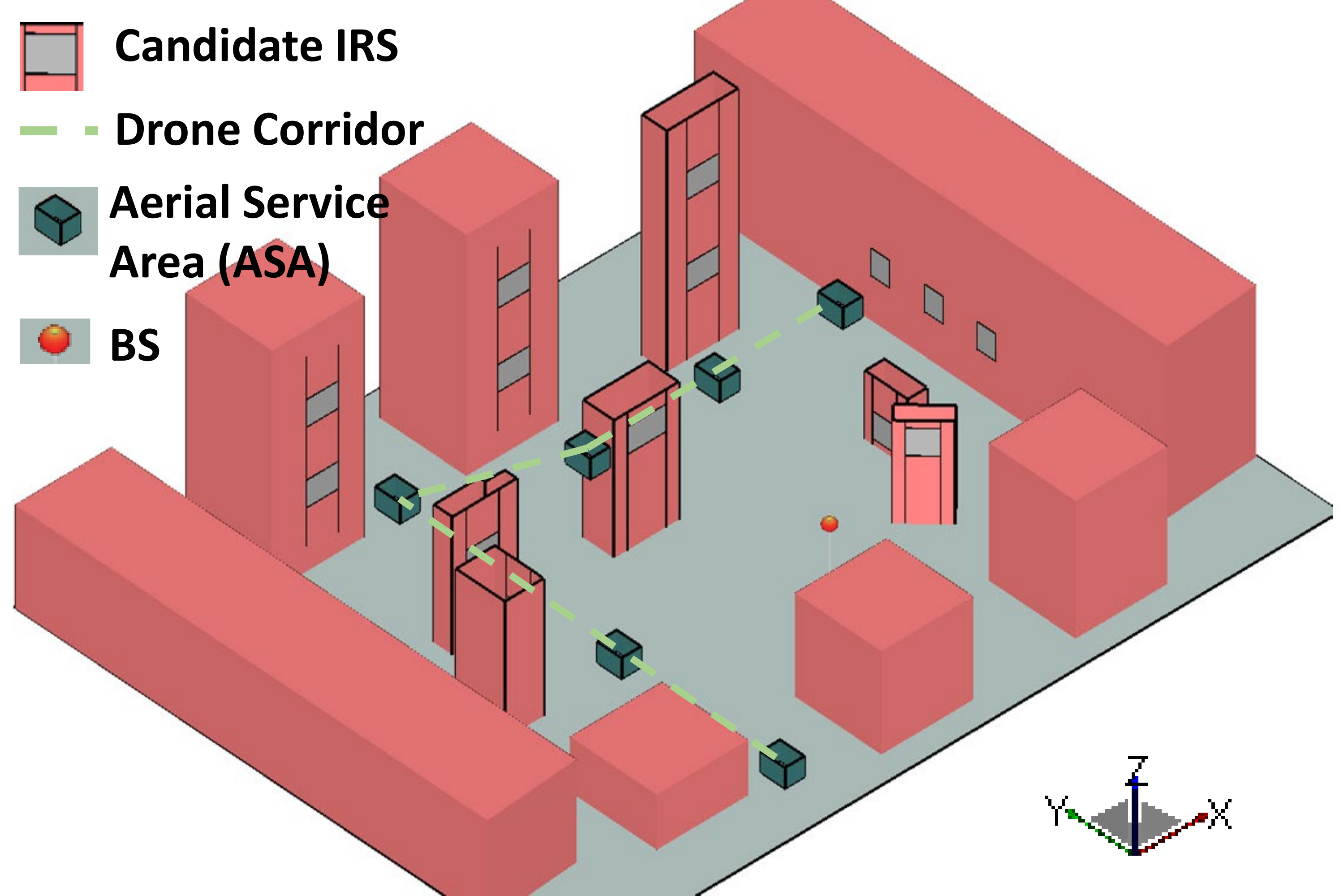}}
\caption{Simulation setup illustrating candidate IRS locations, drone corridor comprised of consecutive ASAs, and BS location in an urban area.} 

\label{Fig:simulation_setup}
\vspace{-3mm}
\end{figure}

The goal of this study, building on the system model presented earlier, is to find the optimal locations of IRSs among a given set of candidate locations, 
in order to maximize the average channel rate at a specific drone corridor while considering certain design criteria~(see Fig.~\ref{Fig:simulation_setup}). The drone corridor is represented by a sequence of uniformly spaced ASAs. Assigning an IRS to a specific ASA requires adjusting the IRS phase map such that the reflected signal from the IRS is directed towards that ASA. The optimization problem's design constraints include: 1) each ASA being served by a single IRS; 2) each IRS serving a single ASA; 3) constraint on the amount of energy that an IRS acquires; and 4) size limitation of IRSs subject to the far-field constraint. We will capture all these limitations by formulating the problem as a binary integer linear programming (BILP) problem.

\begin{figure}
\centerline{\includegraphics[width=0.8\columnwidth]{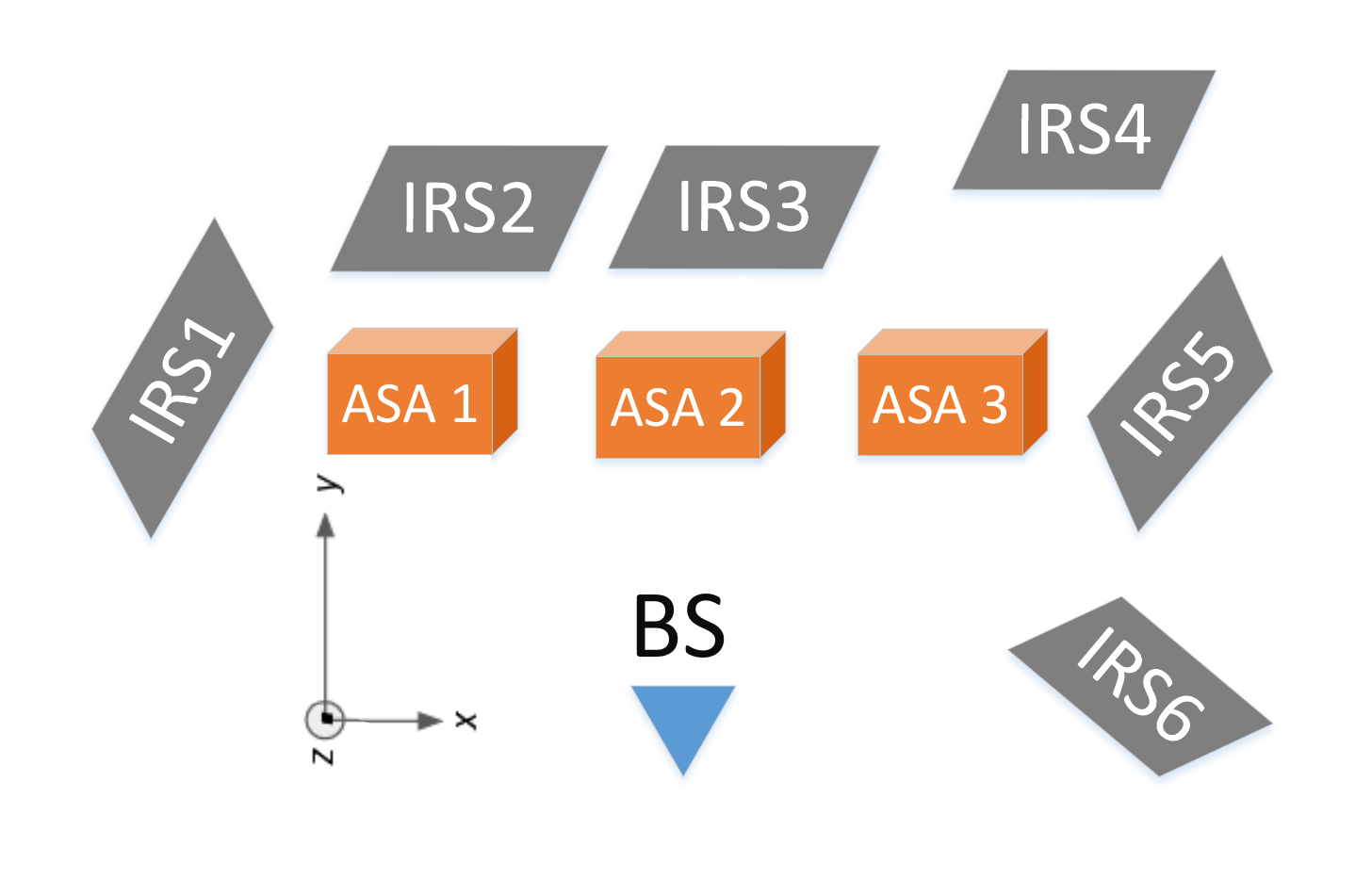}}
\vspace{-5mm}
\caption{A sample scenario consisting of a single BS, 3 ASAs, and 6 IRSs~(grey quadrilaterals) in the universal candidate IRS list (locations in Table~\ref{tab:parameters_list}).} \vspace{-5mm}

\label{Fig:2D_setup}
\end{figure}

The candidate IRS set should be created depending on the environment. For a given an environment, the possible IRS locations should be expressed in terms of reflector center and surface normal representing the facing of the surface. The BS and the ASA locations should be defined before beginning the optimization process. A sample layout in 2D is given in Fig.~\ref{Fig:2D_setup}.

The first step in the process is applying maximum incident angle limitation to create a feasible candidate IRS list for each ASA. The incident angle of the incoming ray from the transmitter on the IRS should be less than a threshold value, $\Phi_c^{\rm i}$. Intuitively, from \eqref{Eq_cosine}, the amount of energy that can be steered by the IRS is proportional with the $\cos^2(\cdot)$ of the incident angle. In this regard, $\Phi_c^{\rm i}$ is a design parameter to narrow down or broaden the candidate IRS list that can serve to a specific ASA. The initial universal set is reduced by eliminating IRSs that do not meet $\phi_m^{\rm tx}<\Phi_c^{\rm i}$ condition, where $\phi_m^{\rm tx}$ is the angle of incidence on the $m^{\text{th}}$ IRS. 

On the reflection side, another limit should be applied, i.e., the departure angle of the outgoing ray from the IRS to the receiver should be less than another threshold value, $\Phi_c^{\rm r}$. This limitation is based on the physical optics approximations used to derive the path loss models. It is seen in Fig.~\ref{Fig:2D_setup} that IRS5 cannot serve any ASA, because of its orientation, whereas IRS6 is dropped from ASA2 and ASA3 candidate IRS lists due to the same reason.  Broadening the list may yield a higher average rate in the drone corridor as the optimization algorithm has wider set of choices at the cost of increased computation time for finding optimal IRS-ASA assignments. The resulting set after applying these two conditions is expressed as:
\begin{align}
     \mathcal{L}^{m, u, t}_{\Phi_c^{\rm i}, \Phi_c^{\rm r}} \in \Big\{(x_m,y_m,z_m, \hat{\textbf{n}}_{m})\quad &|\forall m \in [M], \\
     &\forall \phi_m^{\rm tx} < \Phi_c^{\rm i},\forall \phi_m^{\rm rx} < \Phi_c^{\rm r} \nonumber\Big\},
\end{align}
where $[M]$ represents the set of candidate IRSs. We will refer this set as $\mathcal{L}^{m, u}$ for simplicity.
The optimization problem for finding the IRS-ASA matches can then be formulated as: 
\begin{align}\label{optimization_procedure}
    \max_{\alpha_u^{\mu}}\quad  &\frac{1}{\xi}\sum_u \sum_{\mu}\alpha_u^\mu \text{R}_{u,\mu}\nonumber\\
    \textrm{s.t.} \quad 
    c_0: \quad & \mu \subseteq [M]\nonumber,\\
    c_1:\quad &\alpha_u^m \in \{0,1\} \quad \forall m\in [M], \forall u\in [U],\nonumber\\
    c_2:\quad & \sum_{u=1}^{U} \alpha_u^m \leq 1,\quad \forall m\in \mathcal{L}^{\text{m, u}},\\
    c_3:\quad & \sum_{\mathcal{L}^{\text{m, u}}}\alpha_u^m \leq s_{\text{max}},\quad \forall u\in [U]\nonumber,\\
    c_4:\quad & N_m^{u} =\text{min}\left(\Big{\lfloor} \frac{d^m_{\text{ru}} }{4d^{m}_{\lambda}} \Big{\rfloor}, N_{\text{max}}\right) \forall m\in \mathcal{L}^{\text{m, u}}, \forall u\in [U]. \nonumber
\end{align}
Notation used above is explained further below while describing the objective function and the constraints. 
\begin{itemize}[leftmargin=*]
\item \textit{\textbf{Objective Function:}} The objective of the process is to find the IRS-ASA pairs that maximizes the average rates throughout the drone corridor, where $\xi$ is the cardinality of ASA set, $[U]$, and $\text{R}_{u,\mu}$ is the channel rate from \eqref{Eq_ChanRate}, achieved when all the IRSs in subset-$\mu$ serve the $u^{\text{th}}$ ASA.
\item \textit{\textbf{Constraint 0:}} $\mu$ represents different combinations of IRSs that are chosen to serve an ASA, and it should be a subset of the universal IRS candidate set, $[M]$. 
\item \textit{\textbf{Constraint 1:}} $\alpha_u^m \in \{0,1\}$ is 
1 when $m^{\text{th}}$ reflector is assigned to serve $u^{\text{th}}$ ASA, and $0$ otherwise.

\item \textit{\textbf{Constraint 2:}} Any IRS can be assigned only to a single ASA as 
we consider passive IRSs. 
Passive IRSs are affordable, and less complicated as no control mechanism is required. The sum of $\alpha_u^m$ over $u$ hence should not exceed 1. 

\item \textit{\textbf{Constraint 3:}} The number of IRSs that are assigned to a certain ASA is limited by $s_{\text{max}}$. In theory, it is possible to increase the capacity of the direct channel by $s_{\text{max}}$ folds or even further with high gains if $s_{\text{max}} \leq \text{min}(N_{\rm tx}, N_{\rm rx})$. A higher  $s_{\text{max}}$ yields  increased computational complexity; however, this is not a concern as the optimization process runs offline and the IRSs are placed only once. 

\item \textit{\textbf{Constraint 4:}} Even though the high path loss in mmWave frequencies impel the designers to deploy larger reflectors for higher reflection gains, far-field approximations constitute a natural barrier. In a specific design, the size of an IRS can be the minimum of: 1) the maximum size that is feasible due to physical constraints, and 2) the size of the reflector that the ASA served by that IRS resides at the border of the IRS's Fresnel zone, which is $4a^2/\lambda_c$, where $a$ is the edge dimension of a square reflector. This limit has been expressed in terms of maximum allowed unit cells, $N_{\text{max}}$, assuming a square shaped reflector in (\ref{optimization_procedure}).

\end{itemize}

The BILP problem given in (\ref{optimization_procedure}) is an NP-hard resource allocation problem and can be solved using GUROBIPY library in Python that aims to search for the global optimum.

\section{Numerical Results}\label{section:simulation results}

In this section, we provide simulation results using the proposed BILP-based approach in~\eqref{optimization_procedure}, to find optimum IRS locations for a given BS, three ASA points, and six candidate reflector locations. Simulation parameters are given in Fig.~\ref{Fig:2D_setup} and Table~\ref{tab:parameters_list}. 
The IRS locations are chosen to have heights lower than the ASAs and the BS in compliance with an urban setting where billboards or walls of the buildings can be utilized as reflectors. We consider a $2\times2$ MIMO scheme and three different central frequencies, i.e., $f_{\rm c} =$~ 6, 28, 60~GHz, and three different ASA separations, i.e., $D =$~20, 30, 40~m. The expression given in~\eqref{Eq_ChanRate} are used for the rate calculations. We will explain our findings in subsequent subsections. 

\begin{table}[t]
\renewcommand\arraystretch{1.3}
\caption{Simulation Parameters.}\vspace{-2mm}
\label{tab:parameters_list}
\begin{center}
\begin{tabular}{p{3cm}|p{4.5cm}}
\hline
Parameter & Value \\
\hline
\hline
Center frequencies & $f_c = $ 6, 28, 60~GHz \\
\hline
IRS element spacing & $d^{m}_{\lambda} = 0.25~\lambda$\\
\hline
BS center location 
&(0,0,20)~m\\
\hline
ASAs center locations & ($-D$,20,30) - (0,20,30) - ($D$,20,30)~m\\
\hline
ASA separation & $D = $ 20, 30, 40~m\\
\hline
IRS Locations & (-20,20,30)~(-25,30,10)~(0,30,10) (25,50,10)~(30,20,1)~(25,10,10)~m\\

\hline
Receiver noise power&$\sigma^2 = -$94~dBm\\
\hline
Total power&$P_{\text{tot}}$ = 10~dBm\\
\hline
Antenna gains&$G_{\rm tx} = G_{\rm rx} = 3$~dBi\\
\hline
Max serving IRS & $s_{\text{max}} = 1$\\
\hline
Max allowed incident/reflection angles & $\Phi_c^i = \Phi_c^r = 80$~deg\\
\hline
Max allowed reflector size & 0.25~$\text{m}^2$\\
\cline{1-2}
\end{tabular}
\end{center}
\vspace{-2mm}
\end{table}

\subsection{Frequency Dependency of ASA-IRS Association}

\begin{figure*}
\centering\vspace{1mm}
\captionsetup[subfigure]{labelformat=parens}
\subfloat[]{\includegraphics[width=.75\linewidth]{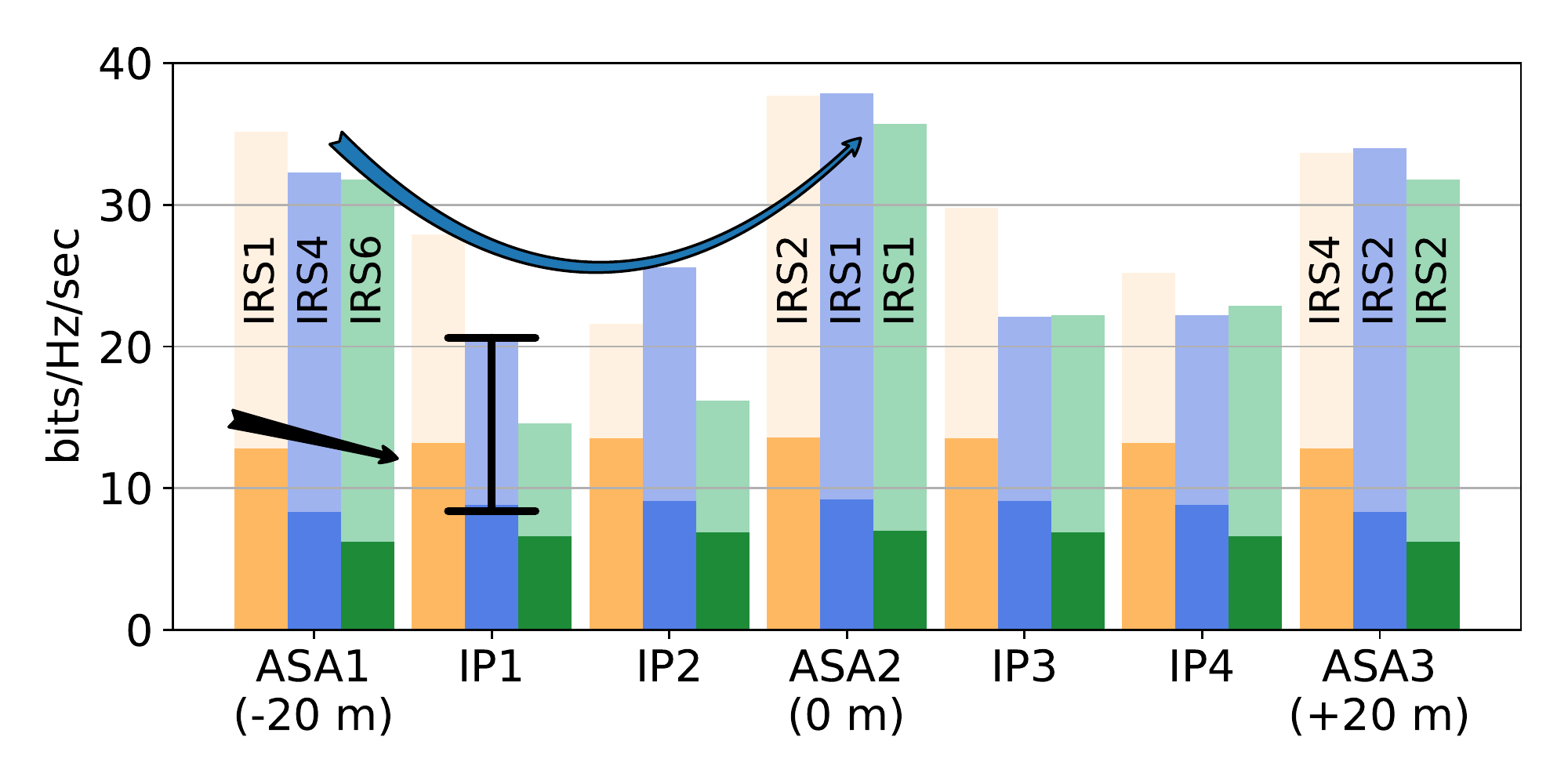}}\vspace{-3mm}
\newline
\subfloat[]{\includegraphics[width=.75\linewidth]{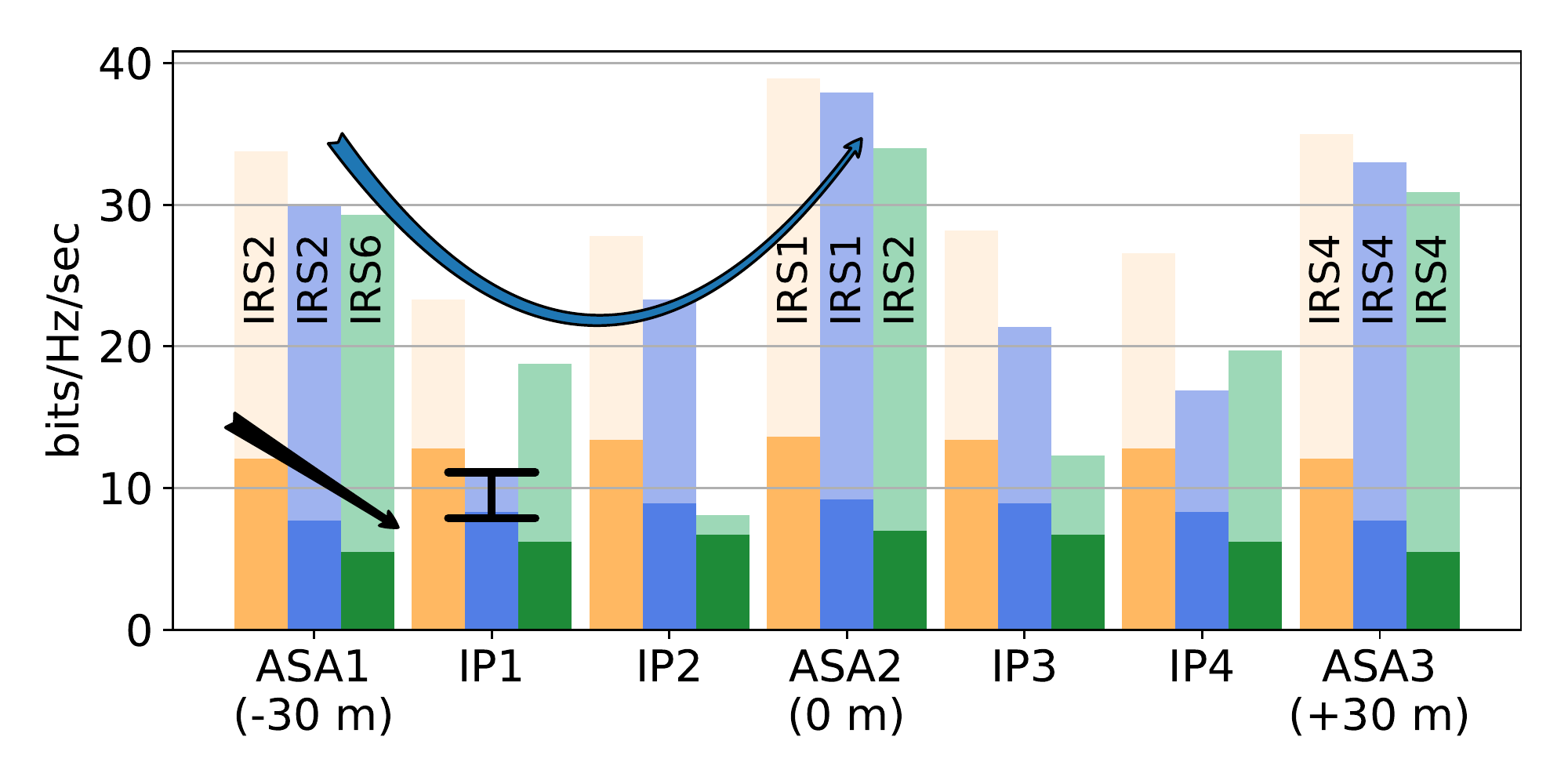}}
\newline
\subfloat[]{\includegraphics[width=.75\linewidth]{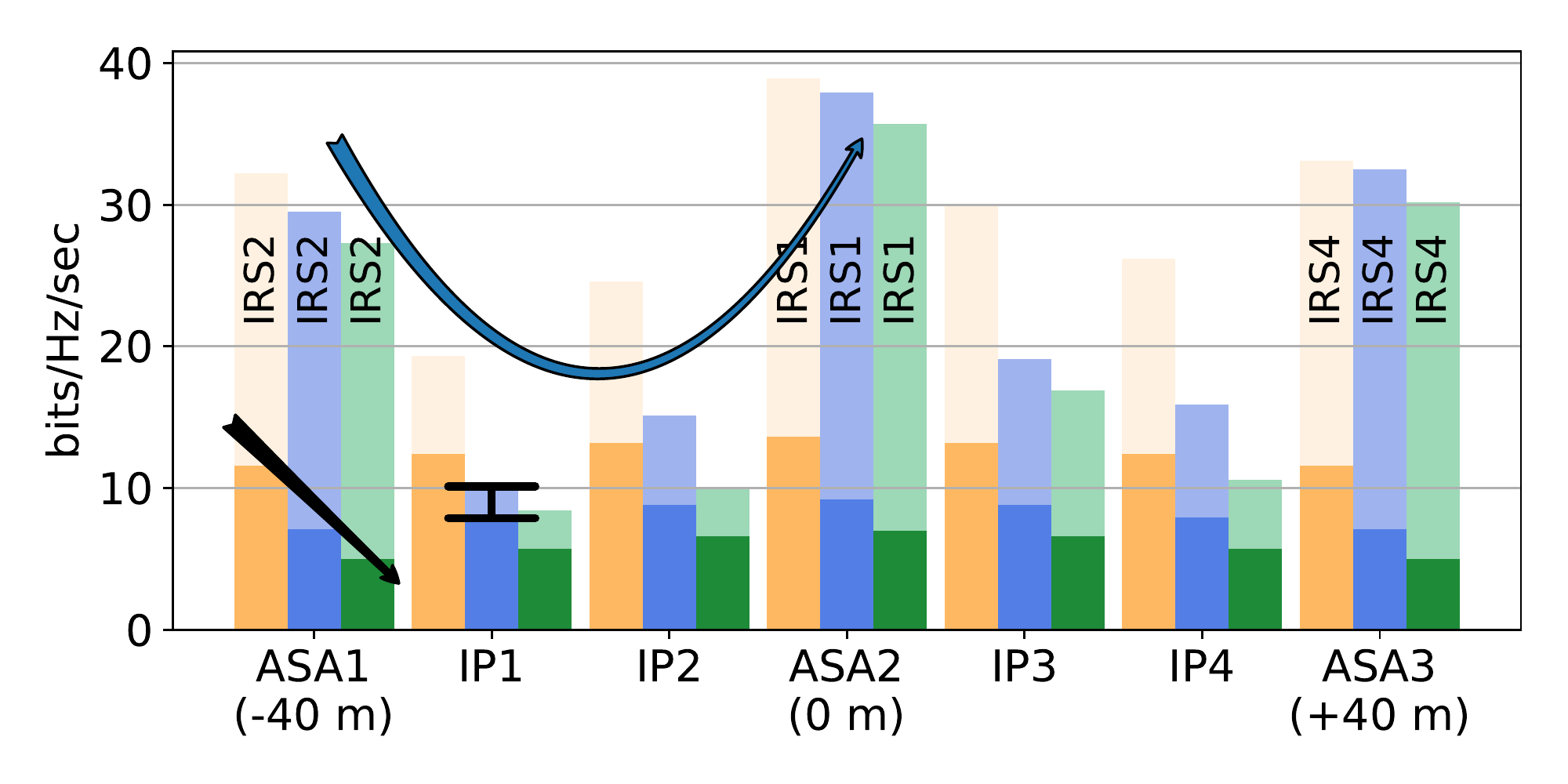}}
\captionsetup[subfigure]{labelformat=empty}
\newline
\subfloat[]{\includegraphics[width=.75\linewidth]{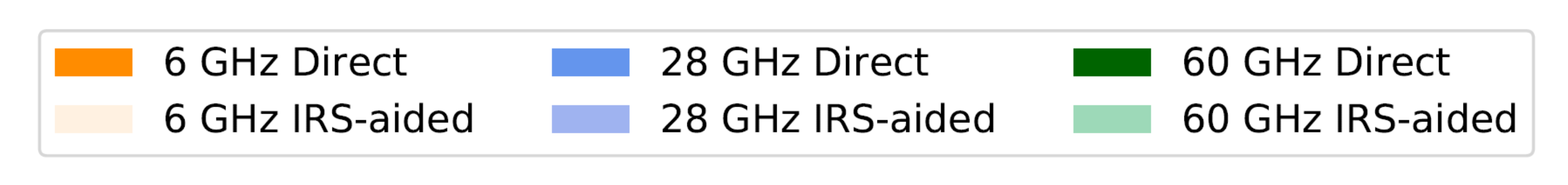}}
\vspace{-4mm}
\caption{Achieved channel rates for different ASA separations:  a) $D=20$~m, b) $D=30$~m, c) $D=40$~m. Direct channel and IRS-aided channel rates, and the associated IRS are given within the same bar for different frequency bands.
As the separation increases, interim point~(IP) rates reduce more. 
} 
\label{Fig:barplot}
\vspace{-4mm}
\end{figure*}

First, the corresponding rates are calculated for all IRS-ASA pairs assuming that the reflector's phase distribution is adjusted to serve that specific ASA, and a database is created. Then the BILP algorithm defines the IRS-ASA matches assuring the maximum average rate over the ASA. The optimized IRS-ASA pairs that give the maximum average rates in the drone corridor are shown in Fig.~\ref{Fig:barplot}. It is observed that the best IRS-ASA matches change by frequency. This is because the path loss behaviour and the Fresnel zones of the IRSs are frequency dependent. For example, IRS~4 is the farthest reflector, and it qualifies to serve an ASA for 6~GHz and 28~GHz bands; however, path loss becomes too high at 60~GHz so that IRS~4 no longer provides adequate rates. Even though IRS~3 is the closest to all three ASAs, due to the lack of angular separation of received rays by ASAs when IRS~3 is in use, channels with IRS~3 give quite low rates~\cite{tse}. Another observation for 20~m separation is that, IRS~1 is matched with ASA~1 in 6~GHz (the closest), however, IRS~4 is matched with ASA~1 in 28~GHz (the farthest). The reason for that is, when the frequency increases, the Fresnel zone of the IRS also widens, and consequently, being close to the receiver becomes no longer advantageous as the reflector size needs to be smaller to keep the ASA outside of the near field.

\subsection{Effect of IRSs on Path Loss}
The achieved channel rates are given for three different frequency bands at three ASAs in Fig.~\ref{Fig:barplot}, together with the direct channel rates. Simulation results show that IRS-aided channels have significantly higher rates; for example, the rate for ASA1 at $20$~m ASA separation increases from 12.8 bits/Hz/sec to 35.2 bits/Hz/sec at $6$~GHz. 

Also note that as the frequency increases, all rates decrease due to increased path loss; however, the IRS-aided channels suffer less than direct channels. Maximum average rates with $D=20$~m ASA separation for $f=$~6, 28, 60~GHz bands are calculated as 35.5, 34.7, 33.1 bits/s/Hz, respectively, whereas, direct channel rates are found to be 13.1, 8.6, 6.5 bits/s/Hz. The IRS-aided channels rates are close to each other even when the frequency increases. The first reason for this behavior is that, a maximum allowed reflector size is used in the simulations, which is given in Table~\ref{tab:parameters_list} as 0.25~m$^2$. For 6~GHz band, this IRS size constraint predominates the far-field constraint, and hence all the IRSs have this maximum allowed size even though a larger IRS would ideally be required. In case that this cap is removed for 6~GHz frequency band, only the far-field constraint remains, and subsequently, best IRS-ASA matches change after the optimization process. With this new assumption, instead of rates and matchings in Fig.~\ref{Fig:barplot}, we would see ASA~1 matches with IRS~6, ASA~2 matches with IRS~1, and ASA~3 matches with IRS~2. Maximum average rate throughout the drone corridor becomes 39.7 bits/s/Hz with average surface area of 0.56~m$^2$. 

The underlying second reason can be better explained rewriting \eqref{Eq_cosine} in terms of the reflector area, $A$, which is defined as $\min(0.25{\rm m}^2, \lambda d_{\rm ru}/4)$ where the second term represents the surface area that ensures the ASA outside of the IRS's Fresnel zone. For mmWave frequencies, the far-field constraint predominates the reflector size constraint: $\beta_c \propto \big(\frac{A}{d_{\rm br} d_{\rm ru}}\big)^2 \propto\big(\frac{\lambda}{4d_{\rm br}}\big)^2$. The reflector index, $m$, and some additional terms are omitted for lucidity.
This expression shows that the introduction of an IRS compensates for the effect of the path loss that comes from the distance between the IRS and the ASA, hence, the increased channel rates. Note that, one component of the compound channel, given in~\eqref{compound_channel}, is the direct channel and the direct channel suffers from the path loss as usual. 

A less influential reason for the channel rates to be close to each other is that, as the frequency band changes, the best IRS-ASA matches also change, which yields to less worse channel rates, i.e., for $D=20$~m separation, in case that the same best IRS-ASA matches of the 28~GHz band are used for the 60~GHz band, the maximum average rate would be 32.5~bits/s/Hz instead of 33.1~bits/s/Hz.

\subsection{Effect of ASA Separation on MIMO Rates}
Fig~\ref{Fig:barplot} shows that, with increased separation between ASAs, the rates calculated for ASA~1 and ASA~3 decrease due to the increased distances and due to the fact that changing the ASA locations eliminates some IRSs from the ASA candidate lists or some IRSs become less useful~(e.g. at $D=30$~m, IRS~1 is no longer an option for ASA~1, and at $D=40$~m, ASA~3 is further away from IRS~1 and IRS~2, and IRS~4 becomes a better option). Hence, overall average achieved rates reduce.

The IRSs are optimized for serving ASAs, however one still expects the reflectors to perform well in between the ASAs, as long as the separation is not high. In this respect, we also studied the performance in the points in between the ASAs in Fig.~\ref{Fig:barplot}, the interim points, when the IRSs are optimized for the ASAs, for different separations of the ASAs. We defined the interim points for each segment (i.e., ASA1-ASA2 segment) as the two points that divides the segment into three equal subsegments. The same scenario described in Fig.~\ref{Fig:2D_setup} is assumed. In these simulations, the location of ASA~2 is kept fixed and, ASA~1 and ASA~3 are moved further away. Separation distances of $D=20,30,40$~m are investigated. 
In general, the calculated rates in Fig.~\ref{Fig:barplot} are closer to the rates at the nearest ASA, however, at some instances, the rates drop significantly. The reason for that is, at those specific points, interim points coincide with the reflector pattern nulls, and consequently, achieved rates decrease drastically. Note that there is a trade-off between the separation distance and average achievable rate throughout the corridor where the interim points are included. One should take the separation into account while designing a drone corridor in case there is a minimum rate requirement throughout the corridor.

\section{Conclusion and Future Work}
\label{conc}

In this letter, we studied the optimal placement of IRSs to maximize the average channel capacity throughout predefined service areas. For a given BS, a set of ASAs (representing a drone corridor), and a set of candidate reflector positions, the proposed optimization approach associates the IRSs with the ASAs so that the average rate with spatial multiplexing is maximized in the drone corridor.
Our simulation results show that the average capacity of the channel in a $2\times2$ MIMO setting increased from 13.1 to 35.5 bits/s/Hz at 6~GHz, from 8.6  to 34.3 bits/s/Hz at 28~GHz, and from 6.5  to 33.1 bits/s/Hz at 60~GHz when the IRSs are used.  

\vspace{-3mm}
\bibliographystyle{IEEEtran}
\bibliography{IEEEabrv,references}

\end{document}